\documentclass[prd,twocolumn,nofootinbib]{revtex4}
\usepackage{bm}
\usepackage{amssymb}
\usepackage{graphicx}
\usepackage{epstopdf}
\usepackage{mathrsfs}
\usepackage{amsmath}

\usepackage[utf8]{inputenc}
\usepackage[english]{babel}
\usepackage{xcolor}

\begin{document}
\title{(Transient) Scalar Hair for (Nearly) Extreme Black Holes} 
\author{Lior M.~Burko$^1$, Gaurav Khanna$^2$, and Subir Sabharwal$^3$}
\affiliation{
$^1$ Theiss Research, La Jolla, California 92037, USA \\ 
$^2$ Department of Physics, University of Massachusetts, Dartmouth, Massachusetts  02747, USA\\
$^3$ Eastmore Group, 40 Wall Street, Suite 1700, New York NY 10005, USA}
\date{June 4, 2019}
\begin{abstract} 

It has been shown recently that extreme Reissner-Nordstr\"{o}m black holes perturbed by a minimally coupled, free, massless scalar field have permanent scalar hair. The hair - a conserved charge calculated at the black hole's event horizon - can be measured by a certain expression at future null infinity: the latter approaches the hair inversely in time.  We generalize this newly discovered hair also for extreme Kerr black holes. We study the behavior of nearly extreme black hole hair and its measurement at future null infinity as a transient phenomenon. For nearly extreme black holes the measurement at future null infinity of the length of the newly grown hair decreases  quadratically in time at intermediate times until its length becomes short and the rate at which the length shortens further slows down. Eventually, the nearly extreme BH becomes bald again like non-extreme BHs.

\end{abstract}
\maketitle

\section{Introduction}

Scalar fields, which are ubiquitous in theoretical physics (e.g, the Higgs field) and in astrophysics (e.g., the inflaton, certain dark matter and dark energy models), have been proposed as candidates for black hole (BH) hair \cite{herdeiro}, in possible violation of the no-hair conjecture. The latter states that all BH solutions of the Einstein-Maxwell equations of general relativity can be completely characterized by three and only three externally observable classical parameters, specifically the BH's mass $M$, charge $q$, and spin angular momentum $a$.  Bekenstein provided a proof for the nonexistence of scalar hair given a set of assumptions \cite{bekenstein1,bekenstein2,bekenstein3}. A number of scalar field hair models have been found, where one or more of the assumptions underlying Bekenstein's theorem are violated. Those include scalar fields with non-strictly positive potentials, scalar fields which are non-canonical or non-minimally coupled to gravity, bound states of ``bald" BHs and solitons \cite{herdeiro}, or in spacetimes with more than four dimensions \cite{dimension}. Also, non-scalar field hair models have been suggested, including non-abelian Yang-Mills \cite{ym-hair} or Proca fields \cite{proca}. In all these examples it is the field itself that constitutes the BH's hair. In addition, when quantum mechanical effects are included, BHs can carry quantum numbers \cite{coleman} and have soft hair \cite{hawking}. 

More recently, a different kind of scalar hair for extreme Reissner-Nordstr\"{o}m (ERN) BHs was found by Angelopoulos, Aretakis, and Gajic \cite{aretakis} (AAG), where a certain quantity $s[\psi]$  evaluated at future null infinity (${\mathscr{I}^+}$) (``measurement at ${\mathscr{I}^+}$ of AAG hair") equals a non-vanishing quantity $H[\psi]$ (``Aretakis charge," ``AAG hair," ``horizon integral") calculated on the BH's event horizon (EH), but vanishes if the BH is non-extreme. Since $H[\psi]$ is a conserved charge for ERN \cite{aretakis1}, it would naturally be related with a candidate for BH hair. Indeed, in \cite{aretakis} it was shown that $s[\psi]$ equals $H[\psi]$. The AAG hair may be construed as a different class of BH hair than the types of hair discussed above, as it is made of minimally-coupled, free, massless scalar field. However, it is not the scalar field itself which constitutes the measurement at ${\mathscr{I}^+}$ of the AAG hair, but a functional of the scalar field $\psi$ which is calculated by adding two terms evaluated at ${\mathscr{I}^+}$, an ``asymptotic term" $s_{\rm I}[\psi]$ and a ``global term", $s_{\rm II}[\psi]$:
\begin{equation}\label{s}
s[\psi]:=\left.\frac{1}{4M}\lim_{u\to\infty}u^2\cdot(r\psi)\right.+\left.\frac{1}{8\pi}\int_{{{\mathscr{I}^+}}\cap\{u\ge 0\}}(r\psi)\right.\,d\Omega\,du\, ,
\end{equation}
where $\psi$  is evaluated on ${\mathscr{I}^+}$ ($\left.\psi\right|_{{\mathscr{I}^+}}$), and $u$ is retarded time. AAG showed that $s[\psi]=H[\psi]$ for ERN, but $s[\psi]=0$ for non-extreme RN BHs, where
\begin{equation}
H[\psi]:=-\frac{M^2}{4\pi}\int_{\rm EH}\,\partial_r(r\psi)\,d\Omega\, ,
\end{equation}
which is calculated on the BH's EH (``AAG hair"). We evaluate below $s[\psi](u)$ by evaluating $s_{\rm I}[\psi](u)$ [without taking the limit in Eq.~(\ref{s})] and by truncating the integration in $s_{\rm II}[\psi]$ at $u$. We evaluate below $H[\psi](v)$ by integrating separately for each value of advanced time $v$.

In what follows we first verify numerically the occurrence of AAG hair for ERN. We then generalize the AAG hair also for extreme Kerr (EK) BHs. We next consider nearly extreme BHs (NERN or NEK, respectively), and show the AAG hair as a transient behavior, including observational features from far away. 

\section{Numerical Method}

Our numerical simulations begin with writing the 2+1 dimensional scalar wave equation in RN or Kerr space-time backgrounds (Teukolsky equation) for azimuthal ($m=0$) modes in compactified hyperboloidal coordinates, which allow us to access $\mathscr{I}^+$ at a finite radial coordinate~\cite{Zenginoglu:2007jw}. The resulting second-order hyperbolic partial differential equation is then re-written as a coupled system of two first-order hyperbolic equations. We then solve this system by implementing a second-order Richtmeyer-Lax-Wendroff iterative evolution scheme~\cite{Zenginoglu:2011zz, Burko:2016uvr}. The initial data are a ``truncated'' Gaussian (to ensure compact support) with non-zero initial field values on the EH. Specifically, in hyperboloidal coordinates $(\rho,\tau)$ (see \cite{Burko:2016uvr} for definitions), the initially  spherical ($\ell=0$) Gaussian pulse is centered at $\rho=1.0M$ with a width of $0.22M$, so that we have horizon penetrating initial data that lead to $H[\psi]\neq0$ on the initial data surface \cite{aretakis}. (For example, the horizon is at $\rho=0.95M$ for ERN and EK in these coordinates.) The Gaussian is truncated beyond $\rho=8.0M$ and the outer boundary is located at $S=\rho(\mathscr{I}^+)=19.0M$. 

In practice, we approximate $H[\psi](v)$ with $H[\psi](\tau)$. At finite times the difference between $\tau$ and $v$ (see Fig.~1 in \cite{Zenginoglu:2011zz}) is manifested in an apparent variation in $H[\psi](v)$ which is a numerical artifact resulting from this approximation. For that reason, the physically relevant value which we use is $H[\psi](v\gg M)$.

\section{Extreme RN/Kerr: Numerical Tests.}

\begin{figure}
\includegraphics[width=8.5cm]{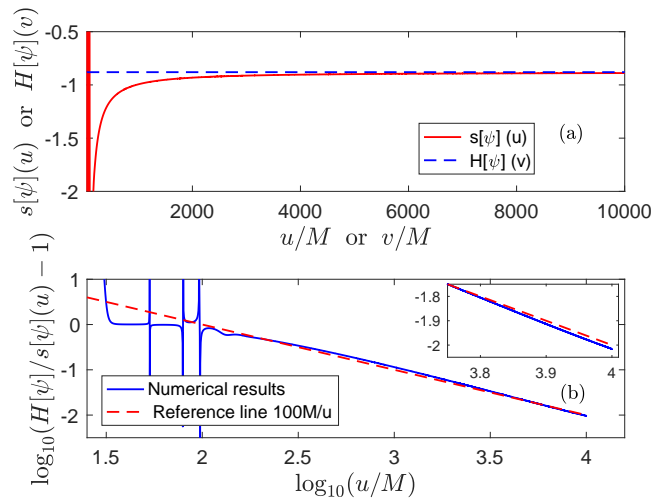}
\caption{Top panel (a): The measurement at ${\mathscr{I}^+}$ of the AAG hair $s[\psi](u)$ (solid curve) and the AAG hair $H[\psi](v)$ (dashed curve), in units of $M^2$, as functions of retarded ($u$) and advanced ($v$) times, respectively, for ERN. Bottom panel (b): The relative difference between them when the EH integral is evaluated for $v\gg M$ (solid curve). The dashed curve is the reference curve $100M/u$.}
\label{RN}
\end{figure}

First, we show in Fig.~\ref{RN} $s[\psi](u)$ and $H[\psi](v)$ as functions of $u$ and $v$,  respectively, for an ERN. Both fields vary as functions of time, although the (unphysical) changes in $H[\psi](v)$ are not visible on the scale of this figure. 
Figure \ref{RN} also shows the relative difference $H[\psi](v\gg M)/s[\psi](u)-1$ as a function of $u$, where $H[\psi](v\gg M)$ approximates $H[\psi](v\to\infty)$. We find that $s[\psi](u)$ approached $H[\psi](v\to\infty)$ for late $u$ as $1/u$ (i.e., $s[\psi](u)\sim H[\psi](v\to\infty)+{\mathscr H}^{\rm RN}[\psi]/u$).  We find for our choice of initial data ${\mathscr H}^{\rm RN}[\psi]\,M^{-3}\sim 100\pm 1$. 

\begin{figure}
\includegraphics[width=8.5cm]{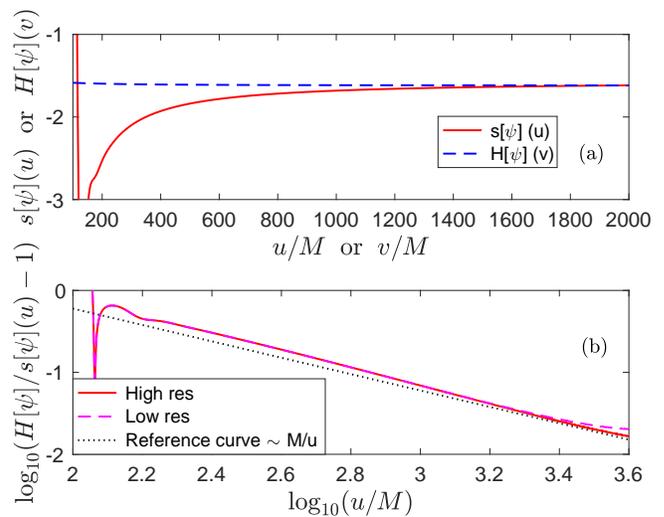}
\caption{Same as Fig.~\ref{RN} for an EK. Notice that the bottom panel (b) shows the results for both low and high grid resolutions.}
\label{Kerr}
\end{figure}

We then apply $s[\psi]$ and $H[\psi]$ also for EK, and present our results in Fig.~\ref{Kerr}. Accurate numerical calculation of $H[\psi]$ is more challenging for EK than for ERN, and requires us to increase the numerical grid density substantially. Figure \ref{Kerr} is the first evidence for AAG hair for EK. We find also for EK that  $s[\psi](u)$ approached $H[\psi](v\to\infty)$ for late $u$ as $1/u$ (i.e., $s[\psi](u)\sim H[\psi](v\to\infty)+{\mathscr H}^{\rm K}[\psi]/u$).  Here, ${\mathscr H}^{\rm K}[\psi]\,M^{-3}\sim 70\pm 1$. 

\section{Nearly and non-extreme RN/Kerr: Numerical Results}

\begin{figure}
\includegraphics[width=8.5cm]{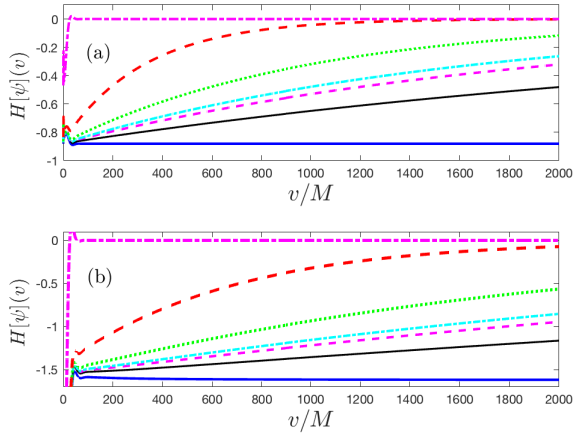}
\caption{The horizon integrals $H[\psi](v)$ (in units of $M^2$) as functions of advanced time $v$ for a number of $1-q/M$ values for RN BHs (top panel, a) and for a number of $1-a/M$ values for Kerr BHs (bottom panel, b). For the top panel from bottom to top, the values are: $1-q/M=0$, $4.5\times 10^{-8}$, $1.25\times 10^{-7}$, $1.8\times 10^{-7}$, $5.0\times 10^{-6}$, $4.5\times 10^{-6}$, and $5.0\times 10^{-5}$. For the bottom panel from bottom to top, the values are: $1-a/M=0$, $4.5\times 10^{-8}$, $1.25\times 10^{-7}$, $1.8\times 10^{-7}$, $5.0\times 10^{-6}$, $4.5\times 10^{-6}$, and $2.0\times 10^{-1}$. }
\label{horizon}
\end{figure}

Next, we consider NERN and NEK. The AAG hair $H[\psi](v)$ is shown in Fig.~\ref{horizon} for a number of $q/M$ and $a/M$ values for RN and Kerr BHs, respectively. For the extreme cases Fig.~\ref{horizon} shows the respective Aretakis charges \cite{aretakis1}. For non-extreme BHs $H[\psi](v)$ attains vanishing values rapidly. For Nearly-extreme BHs $H[\psi](v)$ start at early times with values close to their extreme counterparts, and at late times they approach the non-extreme vanishing values. The closer the BH to extremality, the longer $H[\psi](v)$ takes to get close to zero. 

\begin{figure}
\includegraphics[width=9.5cm]{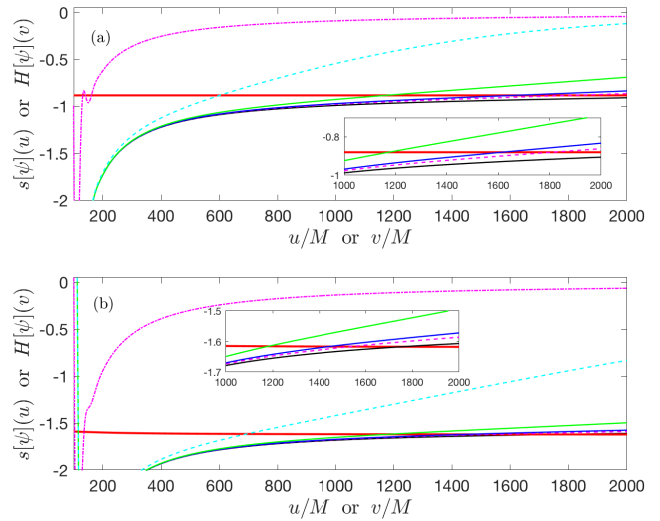}
\caption{The AAG hair $H[\psi](v)$ and its measurement at ${\mathscr{I}^+}$ $s[\psi](u)$ (in units of $M^2$) as functions of $v$ and $u$, respectively, for ERN (top panel, a) and for EK (bottom panel, b), similarly to Figs.~\ref{RN} and \ref{Kerr}, correspondingly. This figure also shows $s[\psi](u)$ for a number of values of $q/M$ and $a/M$, respectively. The insets magnify the late time period of near extremality. The values of $s[\psi](u)$ shown in either panel are the same as in Fig.~\ref{horizon}. We show $H[\psi](v)$ by a nearly horizontal line (variability is unseen on the scale of the figure), at a value of $\sim -0.88$ for the top panel and at a value of $\sim -1.62$ for the bottom.}
\label{hsuv}
\end{figure}

We expect that for nearly-extreme BHs at early times $s[\psi]$ would appear to be similar to that of ERN or EK, respectively, but that at late times it would behave similarly to non-extremal BHs. That is, we expect transient growth of the measurement at ${\mathscr{I}^+}$ of scalar hair for NERN and NEK, after which they would become bald again. Figure \ref{hsuv} shows $s[\psi](u)$ for a number of $a/M$ values for Kerr BHs and for a number of $q/M$ values for RN BHs. The measurement $s[\psi](u)$ approaches a non-zero constant for extreme BHs as $u\to\infty$, whereas $s[\psi](u)\to 0$ for non-extreme BHs. The values of $s[\psi](u)$ for nearly extreme BHs are close at early times to those of their extreme counterparts, but at late times approach those of non-extreme BHs (i.e., vanishing values). The closer the BH is to extremality, the longer it takes to lose its grown hair and achieve baldness. We examine the rate at which this behavior occurs below. 

\begin{figure}
\includegraphics[width=8.5cm]{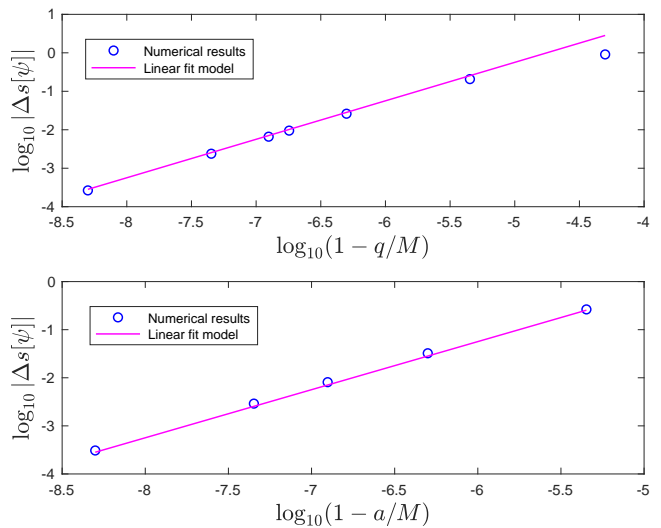}
\caption{The difference between the value of $s[\psi]$ for nearly-extreme BHs and for an extreme BH, $\,\Delta s[\psi]$ (in units of $M^2$), as a function of $1-q/M$ for NERN at $u_*=1000\,M$ (top panel) and as a function of $1-a/M$ for NEK at $u_*=600\, M$ (bottom panel). The numerical data points are shown in circles, and the solid lines are linear best fit lines with slopes $0.997\pm 0.010$ (NERN) and $0.994\pm 0.010$ (NEK).}
\label{deltas}
\end{figure}

The behaviors shown above allow us to distinguish qualitatively between extreme, non-extreme, and nearly-extreme BHs, where the third exhibits transient behaviors between the first and the second. We can obtain quantitative features of the transient nature of $s[\psi](u)$ for nearly-extreme BHs by considering two complementary properties. First, consider a fixed value of retarded time, $u=u_*$, and for a fixed value of $a/M$ or $q/M$ for Kerr or RN BHs respectively, consider for NEK $\,\Delta s[\psi](a/M):=\left. s[\psi] \right|_{u_*} (a/M)-\left. s[\psi]\right|_{u_*}(a/M=1)$, and an analogously defined function of $q/M$ for NERN. In Fig.~\ref{deltas} we plot $\,\Delta s[\psi]$ as a function of $1-a/M$ for NEK and as a function of $1-q/M$ for NERN. For both cases we find that $\,\Delta s[\psi]$ is linear in the distance from extremality.

\begin{figure}
\includegraphics[width=8.5cm]{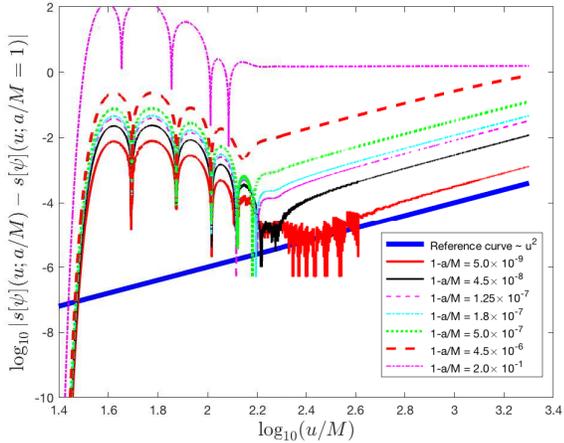}
\caption{The difference  $\,\delta s[\psi]$ for different values of $a/M$ for Kerr BHs as functions of $u$. The reference line (thick solid curve) is $\sim u^2$.}
\label{kerrdeltas}
\end{figure}

\begin{figure}
\includegraphics[width=8.5cm]{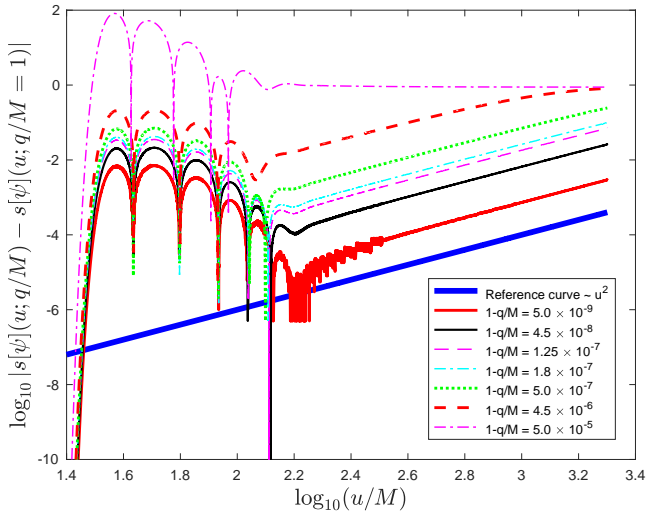}
\caption{Same as Fig.~\ref{kerrdeltas} for values of $q/M$ for RN BHs.}
\label{rndeltas}
\end{figure}

Second, we fix the value of $a/M$ or $q/M$. Define $\,\delta s[\psi](u ; a/M):= s[\psi]  (u ; a/M)-s[\psi]  (u ; a/M=1)$, for NEK and an analogously defined function of $q/M$ for NERN. In Figs.~\ref{kerrdeltas} and \ref{rndeltas} we show $\,\delta s[\psi](u ; a/M)$ as functions of $u$ for NEK and NERN, respectively. The difference between a non-extreme BH and its extreme counterpart is $O(1)$. For nearly extreme BHs the differences $\,\delta s[\psi](u ; a/M)$ or $\,\delta s[\psi](u ; q/M)$ are small at early times (dominated by quasi-normal modes (QNM)), but grow like $u^2$ at intermediate times. At sufficiently late retarded times, which increase with the greater closeness of the BH to extremality, the quadratic growth in retarded time slows down, and  $\,\delta s[\psi]$ approaches its non-extreme BH value asymptotically. For the computations we studied in this work the intermediate regime begins soon after the QNR phase ($\sim 100M$), and then lasts for several hundred to thousands of $M$ depending on $a/M$. 

We can now combine the previous results, and suggest that for NEK
\begin{equation}\label{prop1}
s[\psi]\left(u,\frac{a}{M}\right)=s[\psi]\left(u,\frac{a}{M}=1\right)+{\mathscr{S}}_0^{\rm K}\,u^2\left(1-\frac{a}{M}\right)
\end{equation} 
and for NERN
\begin{equation}\label{prop2}
s[\psi]\left(u,\frac{q}{M}\right)=s[\psi]\left(u,\frac{q}{M}=1\right)+{\mathscr{S}}_0^{\rm RN}\,u^2\left(1-\frac{q}{M}\right)\,,
\end{equation} 
at intermediate times. We find that the dimensionless coefficients ${\mathscr{S}}_0^{\rm K}=0.065\pm 0.001$ and ${\mathscr{S}}_0^{\rm RN}=0.15\pm0.01$ for our choice of initial data. 

\section{Distinguishing extreme, near-extreme and non-extreme RN/Kerr}

This deviation of nearly-extreme BHs from their extremal counterparts allows for their observational identification by distant observers. Specifically, measurements at ${\mathscr{I}^+}$ of a newly perturbed nearly extreme BH shows initial growth of AAG hair. But whereas for EK or ERN where this hair is permanent, for nearly extreme BH the length of the newly grown hair decreases initially quadratically in time until its length becomes short and the rate at which the length shortens further slows down. Eventually the nearly extreme BH becomes bald again like non-extreme BHs. The nearly extreme BH may repeat its hair regrowth attempts when it is perturbed again, but will never succeed for long: It is to eventually lose its regrown hair and become bald again. 

\begin{figure}
\includegraphics[width=9.5cm]{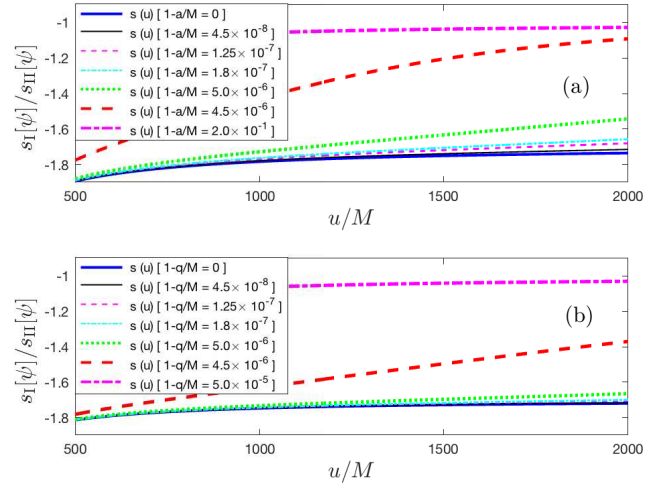}
\caption{The ratio $s_{\rm I}[\psi]/s_{\rm II}[\psi]$ as a function of $u$ for Kerr BHs (top panel, a) and for RN BHs (bottom panel, b). }
\label{relative}
\end{figure}

We can gain additional insight into the transient behavior of NEK and NERN by considering the relative contributions of the two terms in Eq.~(\ref{s}), $s_{\rm I}[\psi]$ and $s_{\rm II}[\psi]$. Figure \ref{relative} shows the ratio $s_{\rm I}[\psi]/s_{\rm II}[\psi]$ for EK and NEK and for ERN and NERN, respectively. For both EK and ERN $s_{\rm I}[\psi]/s_{\rm II}[\psi]\to \sim -1.71$ as $u\to\infty$. For non-extreme BHs $s_{\rm I}[\psi]/s_{\rm II}[\psi]\to -1$ as $u\to\infty$. 
That is, each term in Eq.~(\ref{s}) approaches a non-zero constant for non-extreme BHs, yet their sum vanishes. For nearly extreme BHs Fig.~\ref{relative} shows that at early times the ratio $s_{\rm I}[\psi]/s_{\rm II}[\psi]$ is close to its extreme BH counterpart, but at late times it approaches negative unity, as for non-extreme BHs. We again find that the closer the BH to extremality, the longer it takes the ratio to get close to $-1$. 

Our analysis provides an answer to the question of when a BH is considered nearly extreme. As implied by Figs.~\ref{horizon}, \ref{hsuv}, \ref{rndeltas}, and \ref{relative}, when $1-q/M=5.0\times 10^{-5}$ the transient scalar hair of the BH behaves as for non-extreme BHs. For $1-q/M=4.5\times 10^{-6}$ we already see typical transient behavior, the hallmark of nearly extreme BHs. This effect complements the signature that can be detected by the emission of gravitational waves from a plunge into a nearly extreme BH \cite{burko-kahnna-2016}.

\section*{Acknowledgements}

The authors are indebted to Stefanos Aretakis for stimulating discussions. G.~K.~thanks research support from National Science Foundation (NSF) Grant No. PHY-1701284 and Office of Naval Research/Defense University Research Instrumentation Program (ONR/ DURIP) Grant No.~N00014181255.


\begin{thebibliography}{99}

\bibitem{herdeiro} C.A.R.~Herdeiro and E.~Radu, Int.~J.~Mod.~Phys.~ {\bf D 24}, 1542014 (2015) 
\bibitem{bekenstein1} J. D. Bekenstein, Phys. Rev. Lett. {\bf 28}, 452 (1972)
\bibitem{bekenstein2} J. D. Bekenstein, Phys. Rev. D {\bf 5}, 1239 (1972)
\bibitem{bekenstein3} J. D. Bekenstein, Phys. Rev. D {\bf 5},  2403 (1972)
\bibitem{dimension} C.~Cao, Y.-X.~Chen, and J.-L.~Li, Commun.~Theor.~Phys.~{\bf 53}, 285-290 (2010)
\bibitem{ym-hair} M.S.~Volkov and D.V.~Galtsov, Sov.~J.~Nucl.~Phys. {\bf 51}, 747 (1990) [Yad.~Fiz.~{\bf 51}, 1171 (1990)]; P.~Bizon, Phys.~Rev.~Lett.~{\bf 64}, 2844 (1990)
\bibitem{proca} L.~Heisenberg, R.~Kase, M.~Minamitsuji, and S.~Tsujikawa, Phys.~Rev.~D {\bf 96}, 084049 (2017)
\bibitem{coleman} S.~Coleman, J.~Preskill, and F.~Wilczek, Nuc. Phys. {\bf B 378}, 175-246 (1992)
\bibitem{hawking} S.W.~Hawking, M.J.~Perry, and A.~Strominger, Phys.~Rev.~Lett.~{\bf 116}, 231301 (2016) 
\bibitem{aretakis} Y.~Angelopoulos, S.~Aretakis, and D.~Gajic, Phys.~Rev.~Lett.~{\bf 121}, 131102 (2018)
\bibitem{Zenginoglu:2007jw} A.~Zengino\u{g}lu, Class. Quantum Grav. {P\bf 25} 145002 (2008)
\bibitem{Burko:2016uvr} A.~Zengino\u{g}lu and G.~Khanna, Phys.~Rev.~X {\bf 1}, 021017 (2011)
\bibitem{Zenginoglu:2011zz} L.M.~Burko, G.~Khanna, and A.~Zengino\u{g}lu, Phys.~Rev.~D {\bf 93}, 041501 (2016), [Erratum: Phys.~Rev.~D {\bf 96}, 129903 (2017)]
\bibitem{aretakis1} S.~Aretakis, Commun.~Math.~Phys.~{\bf 307}, 17 (2011);
Annales Henri Poincare {\bf 12}, 1491 (2011) 
\bibitem{burko-kahnna-2016} L.M.~Burko and G.~Khanna, Phys.~Rev.~D {\bf 94}, 084049 (2016) 






\end{thebibliography}
\end{document}